\documentclass[preprint,12pt]{elsarticle}

\usepackage{amssymb}
\usepackage{amsmath}
\usepackage[T1]{fontenc} 
\usepackage[utf8]{inputenc}
\usepackage{float}
\usepackage{lineno}
\usepackage{comment}
\usepackage{xcolor}
\usepackage{makecell}

\usepackage{ucs}
\usepackage[english]{babel}

\usepackage[toc,page]{appendix}
\usepackage{natbib}
\usepackage{comment}
\usepackage[symbol]{footmisc}
\usepackage[colorlinks=true]{hyperref}

\newcommand{\be}{\begin{equation}}
\newcommand{\ee}{\end{equation}}
\newcommand{\ba}{\begin{eqnarray}}
\newcommand{\ea}{\end{eqnarray}}
\newcommand{\epsab}{\varepsilon_{\alpha\beta}}

\newcommand{\epsmm}{\varepsilon_{\mu\mu}}
\newcommand{\epstt}{\varepsilon_{\tau\tau}}

\newcommand{\epsmt}{\varepsilon_{\mu\tau}}

\newcommand{\figref}[1]{Figure~\ref{#1}}

\makeatletter
\renewcommand{\fnum@table}{\textbf{\tablename~\thetable}}
\renewcommand{\fnum@figure}{\textbf{\figurename~\thefigure}}
\makeatother

\makeatletter
\def\ps@pprintTitle{%
	\let\@oddhead\@empty
	\let\@evenhead\@empty
	\def\@oddfoot{\centerline{\thepage}}%
	\let\@evenfoot\@oddfoot}
\makeatother

\begin{document}

\begin{frontmatter}



\title{Search for non-standard neutrino interactions with 10~years of ANTARES data}



\cortext[cor]{corresponding author}

\author[IPHC,UHA]{A.~Albert}
\author[IFIC]{S.~Alves}
\author[UPC]{M.~Andr\'e}
\author[Genova]{M.~Anghinolfi}
\author[Erlangen]{G.~Anton}
\author[UPV]{M.~Ardid}
\author[UPV]{S.~Ardid}
\author[CPPM]{J.-J.~Aubert}
\author[APC]{J.~Aublin}
\author[APC]{B.~Baret}
% ...
\fntext[fn1]{Presently at the University of Cincinnati, Ohio, United States.}
\author[Curtin]{S.J.~Tingay}
\author[IRFU/SPP,APC]{B.~Vallage}
\author[APC,IUF]{V.~Van~Elewyck}
\author[Bologna,Bologna-UNI,APC]{F.~Versari}
\author[LNS]{S.~Viola}
\author[Napoli,Napoli-UNI]{D.~Vivolo}
\author[Bamberg]{J.~Wilms}
\author[Genova]{S.~Zavatarelli}
\author[Roma,Roma-UNI]{A.~Zegarelli}
\author[IFIC]{J.D.~Zornoza}
\author[IFIC]{J.~Z\'u\~{n}iga}

\address[IPHC]{\scriptsize{Universit\'e de Strasbourg, CNRS,  IPHC UMR 7178, F-67000 Strasbourg, France}}
\address[UHA]{\scriptsize Universit\'e de Haute Alsace, F-68100 Mulhouse, France}
\address[IFIC]{\scriptsize{IFIC - Instituto de F\'isica Corpuscular (CSIC - Universitat de Val\`encia) c/ Catedr\'atico Jos\'e Beltr\'an, 2 E-46980 Paterna, Valencia, Spain}}
\address[UPC]{\scriptsize{Technical University of Catalonia, Laboratory of Applied Bioacoustics, Rambla Exposici\'o, 08800 Vilanova i la Geltr\'u, Barcelona, Spain}}
\address[Genova]{\scriptsize{INFN - Sezione di Genova, Via Dodecaneso 33, 16146 Genova, Italy}}
\address[Erlangen]{\scriptsize{Friedrich-Alexander-Universit\"at Erlangen-N\"urnberg, Erlangen Centre for Astroparticle Physics, Erwin-Rommel-Str. 1, 91058 Erlangen, Germany}}
\address[UPV]{\scriptsize{Institut d'Investigaci\'o per a la Gesti\'o Integrada de les Zones Costaneres (IGIC) - Universitat Polit\`ecnica de Val\`encia. C/  Paranimf 1, 46730 Gandia, Spain}}
\address[CPPM]{\scriptsize{Aix Marseille Univ, CNRS/IN2P3, CPPM, Marseille, France}}
\address[APC]{\scriptsize{Universit\'e de Paris, CNRS, Astroparticule et Cosmologie, F-75013 Paris, France}}
% ...
\address[LNS]{\scriptsize{INFN - Laboratori Nazionali del Sud (LNS), Via S. Sofia 62, 95123 Catania, Italy}}
% ...
\address[Roma]{\scriptsize{INFN - Sezione di Roma, P.le Aldo Moro 2, 00185 Roma, Italy}}
\address[Roma-UNI]{\scriptsize{Dipartimento di Fisica dell'Universit\`a La Sapienza, P.le Aldo Moro 2, 00185 Roma, Italy}}
% ...
\address[Bologna]{\scriptsize{INFN - Sezione di Bologna, Viale Berti-Pichat 6/2, 40127 Bologna, Italy}}
% ...
\address[Bologna-UNI]{\scriptsize{Dipartimento di Fisica e Astronomia dell'Universit\`a, Viale Berti Pichat 6/2, 40127 Bologna, Italy}}
% ...
\address[Genova-UNI]{\scriptsize{Dipartimento di Fisica dell'Universit\`a, Via Dodecaneso 33, 16146 Genova, Italy}}
% ...
\address[Curtin]{\scriptsize{International Centre for Radio Astronomy Research - Curtin University, Bentley, WA 6102, Australia}}
% ...
\address[Bamberg]{\scriptsize{Dr. Remeis-Sternwarte and ECAP, Friedrich-Alexander-Universit\"at Erlangen-N\"urnberg,  Sternwartstr. 7, 96049 Bamberg, Germany}}
% ...
\address[IRFU/SPP]{\scriptsize{IRFU, CEA, Universit\'e Paris-Saclay, F-91191 Gif-sur-Yvette, France}}
\address[Napoli]{\scriptsize{INFN - Sezione di Napoli, Via Cintia 80126 Napoli, Italy}}
\address[Roma-Museo]{\scriptsize{Museo Storico della Fisica e Centro Studi e Ricerche Enrico Fermi, Piazza del Viminale 1, 00184, Roma}}
% ...
\address[IUF]{\scriptsize{Institut Universitaire de France, 75005 Paris, France}}
\address[Napoli-UNI]{\scriptsize{Dipartimento di Fisica dell'Universit\`a Federico II di Napoli, Via Cintia 80126, Napoli, Italy}}


\begin{abstract}

Non-standard interactions of neutrinos arising in many theories beyond the Standard Model can significantly alter matter effects in atmospheric neutrino propagation through the Earth. In this paper, a search for deviations from the prediction of the standard 3-flavour atmospheric neutrino oscillations using the data taken by the ANTARES neutrino telescope is presented. Ten years of atmospheric neutrino data collected from 2007 to 2016, with reconstructed energies in the range from $\sim$16~GeV to $100$ GeV, have been analysed. 
A log-likelihood ratio test of the dimensionless coefficients $\varepsilon_{\mu\tau}$ and $\varepsilon_{\tau\tau} - \varepsilon_{\mu\mu}$ does not provide clear evidence of deviations from standard interactions. For normal neutrino mass ordering, the combined fit of both coefficients  yields a value 1.7$\sigma$ away from the null result.  However, the 68\% and 95\% confidence level intervals for $\varepsilon_{\mu\tau}$ and $\varepsilon_{\tau\tau} - \varepsilon_{\mu\mu}$, respectively, contain the null value. Best fit values, one standard deviation errors and bounds at the 90\% confidence level for these coefficients are given for both normal and inverted mass orderings. The constraint on $\varepsilon_{\mu\tau}$ is among the most stringent to date and it further restrains the strength of possible non-standard interactions in the $\mu - \tau$ sector.
\end{abstract}

\end{frontmatter}


\section{Introduction}
\label{sec:intro}

Neutrinos have provided the first hint of physics beyond the Standard Model (BSM) through the discovery of neutrino oscillations~\cite{Ahmad:2002jz, Fukuda:1998mi}, which imply that at least two of the three neutrinos presently known have non-zero masses~(see chapter 14 in ref.~\cite{RPP2020}). The origin and smallness of the neutrino masses compared to the rest of the Standard Model (SM) particles point to new physics at a very high energy scale~\cite{Minkowski:1977sc, Mohapatra} and is at present the subject of intense research. 




Interactions not present in the SM are predicted by a wide variety of BSM theories. Considering the presence of new physics at a high energy scale, the effect at a lower scale can be approached in a model independent way by constructing a sum of operators built from the fields which are relevant at the low energy scale. BSM will then show up at a lower scale via non-renormalisable operators with dimension five or higher~\cite{Weinberg:1979sa, Costa:1985vk}. In the context of neutrino physics, out of all possible neutrino interactions that can be built within this effective theory approach~\cite{Bischer:2019ttk}, a subset called Non-Standard Interactions (NSIs) has attracted particular attention~\cite{Ohlsson:2012kf, Miranda:2015dra}. The NSIs are quantified through dimensionless constants ($\epsab$) that appear in the four-fermion interactions expressed as


\begin{align}
 \mathcal{L}^{\text{CC}}_{\text{NSI}} &=  -2\sqrt{2}G_{F} \, \sum_{ff'X} \varepsilon^{ff' \, X}_{\alpha\beta} \, (\overline{\nu}_{\alpha} \, \gamma^{\mu} P_{L}\, l_{\beta}) \, (\overline{f} \, \gamma_{\mu}P_{X}f') ,\\
\mathcal{L}^{\text{NC}}_{\text{NSI}} & =  -2\sqrt{2}G_{F} \, \sum_{f X} \, \varepsilon^{f \, X}_{\alpha\beta} \,
 (\overline{\nu}_{\alpha} \, \gamma^{\mu}  P_{L}\, \nu_{\beta}) \, (\overline{f} \, \gamma_{\mu}  P_{X}f),
\end{align}
where $G_{F}$ is the Fermi constant, $P_{X}$ (with $X$=$R$ or $L$) denotes the chiral projection operators $P_{R,L}$=$\frac{1}{2}$ 
(1$\pm\gamma^{5})$, $f$ is a first generation SM fermion ($e$, $u$ or $d$-quarks), $f'$ belongs to the same weak doublet as $f$, and $\alpha$ and $\beta$ denote the neutrino flavours: $e$, $\mu$ or $\tau$. The dimensionless coefficients $\varepsilon^{ff' \, X}_{\alpha\beta}$ and $\varepsilon^{f \, X}_{\alpha\beta}$ quantify the strength of NSIs between the neutrinos of flavour $\alpha$ and $\beta$ and the fermion $f \in$ \{$e$, $u$ or $d$\} (for neutral currents) and $f \neq f' \in$ \{$u$, $d$\} (for charged currents). The SM scenario is recovered in the limit $\varepsilon \to$~0.

While charged current (CC) NSIs affect the production and detection processes of neutrino states at scattering experiments~\cite{COHERENT:2017ipa}, neutral current (NC) NSIs would affect the neutrino propagation by coherent forward scattering. In this paper, we constrain NC NSIs that alter the propagation of neutrinos while travelling long distances through the Earth. 

 The possible existence of NSIs can be studied in a variety of experiments~\cite{Miranda:2015dra, Kopp:2007ne, Farzan:2017xzy, Dev:2019anc}. In particular, atmospheric neutrinos provide an excellent opportunity since NSIs will modify the SM potential that describes the matter effects in neutrino flavour oscillations. This will give rise to additional effects that can produce deviations from the expectations for the standard oscillation phenomenon in matter~\cite{Miranda:2015dra, Farzan:2017xzy, Dev:2019anc, Fornengo:2001pm, GonzalezGarcia:2004wg}. High-energy neutrino telescopes are suitable detectors to search for NSI-induced deviations since they are expected to increase with the distance travelled through matter and with energy. Limits on NSIs from atmospheric neutrinos have already been reported by the Super-Kamiokande Collaboration~\cite{Mitsuka:2011ty} or using Super-Kamiokande data~\cite{Fukasawa:2015jaa}, and by the IceCube Collaboration~\cite{Aartsen:2017xtt, IceCubeCollaboration:2021euf, IceCube:2022ubv} or using IceCube data~\cite{Esmaili:2013fva, Salvado:2016uqu, Demidov:2019okm}.

In this paper, a search for NSI-induced deviations from the expected neutrino oscillation process using 10 years of ANTARES data~\cite{Collaboration:2011nsa} is presented.
 
 The article is organised as follows. In section~\ref{sec:NSI}, the standard paradigm of neutrino flavour oscillations both in vacuum and in matter, as well as the deviations that the presence of NSIs would introduce, are summarised. In section~\ref{sec:detector}, the ANTARES neutrino telescope is described. In section~\ref{sec:dataset}, the ANTARES data sample used for this analysis are presented and the Monte Carlo (MC) simulation of neutrino events, their reconstruction and the final event selection is discussed. The details of the analysis are given in section~\ref{sec:analysis} and the results, in section~\ref{sec:results}. Finally, the conclusions of this work are gathered in section~\ref{sec:conclusion}.

\section{Neutrino propagation in matter and NSIs}
\label{sec:NSI}

\textcolor{red}{}
As shown by a series of experiments with solar, atmospheric, accelerator and reactor neutrinos~\cite{RPP2020}, neutrino flavour eigenstates are different from neutrino mass eigenstates. The flavour eigenstates are involved in neutrino production and annihilation in weak interactions, while the mass eigenstates determine the neutrino propagation. Since a neutrino produced with a given flavour evolves according to its superposition in terms of matter eigenstates, it can later interact as a different flavour, giving rise to neutrino flavour oscillations.

Neutrino oscillation probabilities are governed by the Pontecorvo-Maki-Nakawaga-Sakata mixing matrix (PMNS)~\cite{Pontecorvo:1967fh,Maki:1962mu} and the differences of squared masses. In the conventional three-flavour scheme with Dirac neutrinos, the relevant parameters are three mixing angles, $\theta_{12}$, $\theta_{13}$, $\theta_{23}$, two mass-squared differences, $\Delta m^2_{21}$, $\Delta m^2_{31}$, and a CP violating phase, $\delta_{\text{CP}}$. Our present knowledge of these parameters can be summarised with the following ranges (see chapter 14 in ref.~\cite{RPP2020}): 
$\theta_{12} \approx$ 31$^\circ$--36$^\circ$, 
$\theta_{13} \approx$ 8$^\circ$--9$^\circ$, 
$\theta_{23} \approx$ 41$^\circ$--51$^\circ$, 
$\Delta m^2_{12} \approx$ 7--8 $\times$ 10$^{-5}$ eV$^2$, 
$\lvert\Delta m^2_{23}\rvert$ ($\approx \, \lvert\Delta m^2_{13}\rvert) \, \approx$ 2.4--2.6 $\times$ 10$^{-3}$ eV$^2$ and 
$\delta_{\text{CP}} \approx  $ 200$^\circ$ -- 350$^\circ$. The octant of $\theta_{23}$, the sign of $\Delta m^2_{23}$ (the mass ordering) and an accurate value of $\delta_{\rm CP}$ are yet to be determined. A number of neutrino experiments are planned to improve these measurements~\cite{KM3NeT:2021ozk, JUNO:2021vlw, Adrian-Martinez:2016fdl, collaboration2014letter,   abe2011letter, 2015, Akindinov_2019, abi2020deep}.

When neutrinos propagate in matter, their evolution is affected by the coherent forward elastic scattering on medium. The overall effect can be described by effective potentials associated to the CC and NC interactions. In the case of neutrinos travelling through the Earth, the only relevant potential is the one stemming from the electron neutrino components interacting with electrons in matter. The potential is given by $V_{CC} = \sqrt{2} \, G_{F} \, n_{e}$, where 
$n_{e}$ is the electron number density along the neutrino path~\cite{Wolfenstein:1977ue}. The relevant Hamiltonian is

\begin{equation}
H^{3\nu}= \frac{1}{2E_{\nu}} \, U M^2 U{^\dagger} + V_{CC} \, diag(1,\,0,\,0)
\label{standardHam}
\end{equation}

\par\noindent The PMNS mixing matrix, $U$, performs the rotation of the relevant mass matrix $M^2=diag(0, \Delta m^{2}_{21}, \Delta m^{2}_{31})$ in the neutrino flavour space, where $diag$ indicates a diagonal matrix with the specified elements.  

The model-independent NSIs are introduced as new potentials,

\begin{equation}
H^{\rm NSI}=  V_{CC} \, \frac{n_f}{n_e} \, {\bf \varepsilon},
\end{equation}

\par\noindent where $n_f$ is the fermion number density along the neutrino path. $H^{\rm NSI}$ is added to the SM Hamiltonian in Eq.~\ref{standardHam}. In the present work, neutrinos are assumed to interact with down quarks which are roughly three times as abundant as electrons, $n_{f} = n_{d} \approx$ 3 $n_{e}$. The matrix $\varepsilon$ ($\varepsilon_{\alpha \, \beta}$, $\alpha, \, \beta= e, \mu$, $\tau$) gives the strength of NSIs. Its diagonal terms, if different from each other, give rise to the violation of leptonic universality, while the off-diagonal terms induce flavour changing neutral currents, which are highly suppressed in the SM. 


The hermiticity of the hamiltonian matrix reduces the components of  the $\varepsilon$ matrix to nine real parameters, the three real diagonal plus the three complex off-diagonal elements. In addition, in oscillation experiments this matrix can only be determined up to a global multiple of the identity matrix, so only two parameters of the diagonal are independent, which are conventionally taken to be $\varepsilon_{e e} - \varepsilon_{\mu \mu}$ and $\varepsilon_{\tau \tau} - \varepsilon_{\mu \mu}$. 

To reduce the number of parameters to be fitted, some assumptions are customarily made depending on the nature of the experiment.
Since ANTARES detects atmospheric neutrinos and in this study only track-like events are used, we essentially observe charged current interactions of atmospheric muon neutrinos, which are known to oscillate mainly to tau neutrinos\footnote{By "neutrinos" we implicitly mean neutrinos and anti-neutrinos, unless otherwise stated. }.  
Therefore, we are mostly sensitive to the NSI parameters $\varepsilon_{\mu \tau}$ and $\varepsilon_{\tau \tau} - \varepsilon_{\mu \mu}$. In this study, we assume that all other parameters are zero\footnote{A common approach to tackle this large number of parameters is to fit them one-by-one, fixing the rest to zero. This procedure was adopted for instance in ref.~\cite{IceCubeCollaboration:2021euf}.}. 
Furthermore, since our sensitivity to phases is low, we assume $\varepsilon_{\mu \tau}$ to be real, although this implies a loss of generality.

NSI are supposed to manifest themselves as sub-dominant effects on top of the main phenomenon of neutrino oscillations. In our case, the atmospheric neutrino oscillation channel $\nu_{\mu}\rightarrow\nu_{\tau}$ gives rise to the disappearance of muon neutrinos with a probability that depends on their energy and baseline. Therefore, a deficit with respect to the non-oscillation hypothesis 
should be observed in the detector as a function of the neutrino energy and arrival direction.
In the range from $\sim$10~GeV to slightly above $\sim$100~GeV, the detector is capable of estimating the neutrino energy by measuring the muon range~\cite{Albert:2018mnz}. The first oscillation minimum in the $\nu_{\mu}\rightarrow\nu_{\mu}$ channel occurs around 20 GeV, and the $\nu_{\mu}$ deficit is noticeable all the way up to 100 GeV, which enables ANTARES to measure the atmospheric neutrino oscillation parameters (sin$^{2}\theta_{23},\,\Delta m^2_{23}$).

\begin{figure}[htb]
\includegraphics[width=0.5\textwidth]{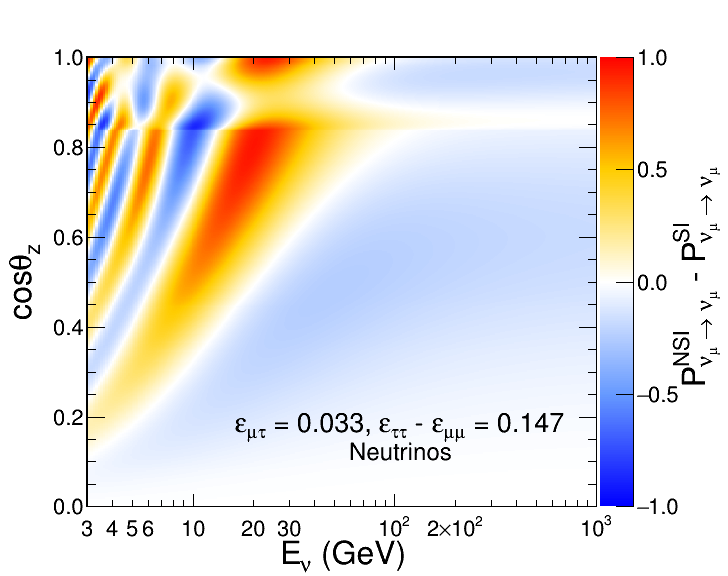}
\includegraphics[width=0.5\textwidth]{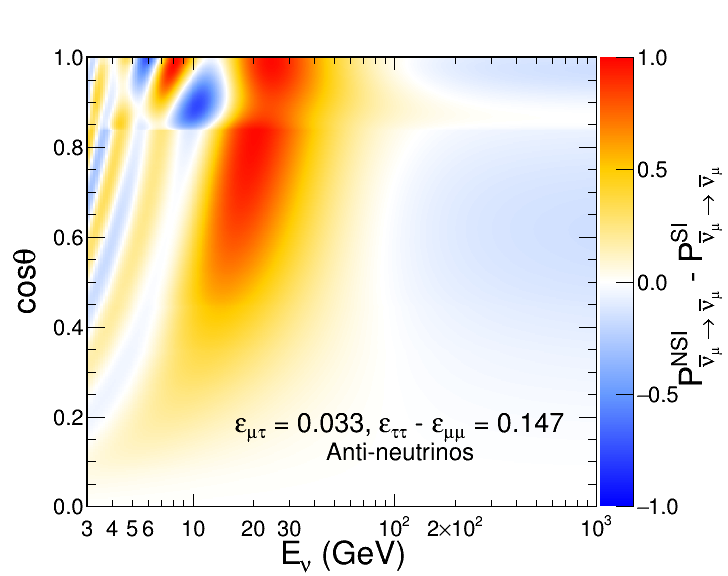}

\caption[]{NSI-induced modifications in $\nu_{\mu}$ (left) and $\bar{\nu}_{\mu}$ (right) disappearance probabilities as a function of the true neutrino energy and cosine of the zenith angle. The NSI test point has been set at $(\varepsilon_{\mu\tau}, \varepsilon_{\tau\tau} - \varepsilon_{\mu\mu}) = (0.033, 0.147)$. Normal mass ordering is assumed.}
\label{fig:numuNSImutau}
\end{figure}

As an example of the effect of NSI, \figref{fig:numuNSImutau} shows the change in the oscillation pattern arising from non-zero values of $\epsmt$ and $\epstt - \epsmm$. The difference of the survival probabilities for the case of NSI  ($\text{P}^{\text{NSI}}_{\nu_{\mu} \rightarrow \nu_{\mu}}$) and of standard oscillations ($\text{P}^{\text{SI}}_{\nu_{\mu} \rightarrow \nu_{\mu}}$) is plotted as a function of the cosine of the zenith angle\footnote{A value cos$\,\theta = 1$, corresponds to vertically upgoing neutrinos traversing the Earth core and passing through its density layers, whereas cos$\,\theta$ = 0 corresponds to horizontally moving neutrinos.}, cos$\,\theta$ and the neutrino energy, $E_{\nu}$, in Figure~\ref{fig:numuNSImutau}. The left plot shows the difference for neutrinos and the right plot for anti-neutrinos. The test values of the oscillation parameters are adopted from the global neutrino oscillation fit results in ref.~\cite{Esteban:2016qun}. Normal mass ordering is assumed. The NSI test point is chosen at $(\varepsilon_{\mu\tau}, \varepsilon_{\tau\tau} -  \varepsilon_{\mu\mu}) = (0.033, 0.147)$\footnote{These values are motivated by the 90\% C.L. limits set by the Super-Kamiokande experiment~\cite{Mitsuka:2011ty}.}.
 Since the NSI effects are not symmetric
for neutrinos and anti-neutrinos, they lead to a small but discernible effect in the combined $\{\nu + \bar{\nu}\}$ event distribution. This is important for the ANTARES detector which measures neutrinos via muons and is charge blind.
It can be noticed that the effect of NSIs becomes prominent for higher energies and vertical upgoing events. 


\section{The ANTARES Detector}
\label{sec:detector}

The ANTARES neutrino telescope~\cite{Collaboration:2011nsa} is located in the Mediterranean Sea, 40\,km off the coast of Toulon, France, at a depth of about 2475\,m. The detector, which was completed in 2008, is composed of 12 detection lines, each one equipped with 25 storeys of 3 optical modules (OMs), except line 12 with only 20 storeys of OMs, for a total of 885 OMs. The horizontal spacing among the lines is $\sim$60\,m, while the vertical spacing between the storeys is 14.5\,m. Each OM hosts a 10-inch diameter photomultiplier tube (PMT) from Hamamatsu~\cite{Amram:2001mi}, whose axis points 45$^{\circ}$ downwards. All signals from the PMTs that pass a threshold of 0.3 photoelectrons are digitised (\textit{hits}) and sent to the shore station. 
The position of the optical modules is determined with an accuracy of $\sim$10 cm~\cite{AdrianMartinez:2012ky} and the overall time calibration is better than 1~ns~\cite{Aguilar:2010aa}. The main sources of optical background registered by the ANTARES PMTs are the Cherenkov light induced by the decay products of the radioactive isotope $^{40}\mathrm{K}$ and the bioluminescence. A detailed description of the ANTARES detector and further information about the data acquisition, trigger and calibration systems can be found in~\cite{Collaboration:2011nsa}.

\section{Data set, simulation, reconstruction and event selection}
\label{sec:dataset}

The data sample used in this work is the same as the one used in the previous ANTARES oscillation analysis~\cite{Albert:2018mnz}. In that study, the data recorded by ANTARES from 2007 to 2016 (both years included), corresponding to a total livetime of 2830 days, was analysed to select a final sample of 7710 reconstructed track-like events. With this sample, a muon neutrino disappearance study was performed which provided a measurement of the oscillation parameters sin$^{2}\theta_{23}$ and $\Delta m^2_{23}$ and established limits on the (3+1) sterile neutrino mixing model. The main features of the simulation, event reconstruction and selection are briefly described in what follows. Further details can be found in refs.~\cite{Albert:2018mnz, ANTARES2020bhr, Ilenia}.

CC $\nu_{\mu}$ interactions in seawater produce muons propagating through the detector that induce the emission of Cherenkov photons. They are identified as track-like events.  All other types of neutrino interactions give rise to Cherenkov photons in the shower-like topology. The event reconstruction and selection used in the analysis have been optimised to select track-like events. On the other hand, $\nu_{e}$ CC interactions and NC interactions of all flavours produce hadronic showers. In the case of $\nu_{e}$ CC interactions, an electromagnetic shower is produced as well. Moreover, $\nu_\tau$ CC events can be produced as the result of $\nu_\mu\rightarrow\nu_\tau$ oscillations with or without muons in the final state. All these events constitute an additional source of background for this study.

Simulated atmospheric neutrinos are generated following the flux calculation of Honda \textit{et~al.}~\cite{Honda:2015fha}. Neutrino interactions surrounding ANTARES and inside the instrumented volume are simulated using the GENHEN software package~\cite{Genhen}. Atmospheric muon events are simulated by MUPAGE~\cite{Carminati:2008qb}. Particle propagation is done using a GEANT-based package~\cite{Agostinelli:2002hh}, which also simulates the Cherenkov light production and propagation taking into account the seawater optical properties~\cite{Aguilar:2004nw}. Optical background from $^{40}$K decays and bioluminescence in water, as measured from counting rates in data, is added on a \textit{run-by-run} basis~\cite{Fusco:2016jvz}, which provides a very accurate simulation of the environmental conditions. The overall geometry of the detector, the angular acceptance and quantum efficiency of the PMTs~\cite{Amram:2001mi}, as well as the response of the electronics is then taken into account to produce simulated digitised signals~\cite{ANT2010elec}.

Two different muon track reconstruction algorithms are used in this analysis~\cite{Aguilar_2011, Reco}. In method $\mathcal{A}$, a hit selection based on time and spatial coincidences of hits is applied and a $\chi^2$-fit is performed in order to extract the best track parameters. Events reconstructed by this method can have a single-line topology, if all the selected hits are recorded in the same detector line, or a multi-line topology, when hits belong to OMs of different lines. Method $\mathcal{B}$ involves a first prefit based on a directional scan of isotropically distributed directions, followed by a final log-likelihood fit of the track parameters using the best directions as starting points.

In the 10--100 GeV energy range, muons in water behave as minimum ionising particles, so that their energy is proportional to their pathlength, with a proportionality constant of 0.24~GeV/m (see chapter 34 in ref.~\cite{RPP2020}). The track length of the muon in the detector, $L_{\mu}$, can be calculated projecting back to the reconstructed track the first and the last selected hits. For single-line events, the track length is estimated from the vertical coordinates of the uppermost and lowermost storeys with selected hits and the reconstructed zenith angle. The reconstructed energy from the muon track length does not take into account the energy of the hadronic shower in the interaction vertex, nor the fact that the muon track might be only partially contained in the detector sensitive volume. Whereas for multi-line events, the threshold energy of the final neutrino sample is about 50 GeV, for nearly-vertical single-line events the energy estimate can be as low as 20 GeV or less, close to the first oscillation minimum. In the 20~GeV energy regime, the median angular resolution is 3$^{\circ}$ for single-line and 0.8$^{\circ}$ for multi-line events and  the energy resolution for muons is found to be (50 $\pm$ 22)\%~\cite{Yanez:2015uta}.

The reconstructed muon energy of fully and partially contained events extends to around 100~GeV, but the true energy of the parent neutrinos goes up to a few TeVs. This broad range of neutrino energy brings enhanced sensitivity to the NSI parameters. 

The events are selected using a quality criterion optimised using Monte Carlo simulated events. This quality criterion is based on the goodness-of-fit provided by each reconstruction algorithm: a reduced-$\chi^2$ for method $\mathcal{A}$ and a log-likelihood for method $\mathcal{B}$. Events reconstructed by any of the two methods that fulfill the corresponding quality criterion and that are upgoing (cos$\,\theta^{\rm reco}$ > 0.15) are kept for the analysis.    

When these criteria are applied to a Monte Carlo sample of atmospheric neutrinos without oscillations plus atmospheric muons, corresponding to the 2830 days of data, a total of 8044 events are selected.
More than 95\% of these selected events are muon neutrinos, the background being mainly composed of atmospheric muons misreconstructed as upgoing. When applied to the data sample, the procedure yields a total of 7710 events, 5632 from method $\mathcal{A}$ (1950 from single-line and 3682 from multi-line) and 2078 from method $\mathcal{B}$, in agreement with the expectations taking into account oscillations~\cite{Albert:2018mnz}. Event samples from both methods, $\mathcal{A}$ and $\mathcal{B}$, are selected for the final analysis.

\section{Analysis}
\label{sec:analysis}

\begin{figure}[ht]
\includegraphics[width=0.5\textwidth]{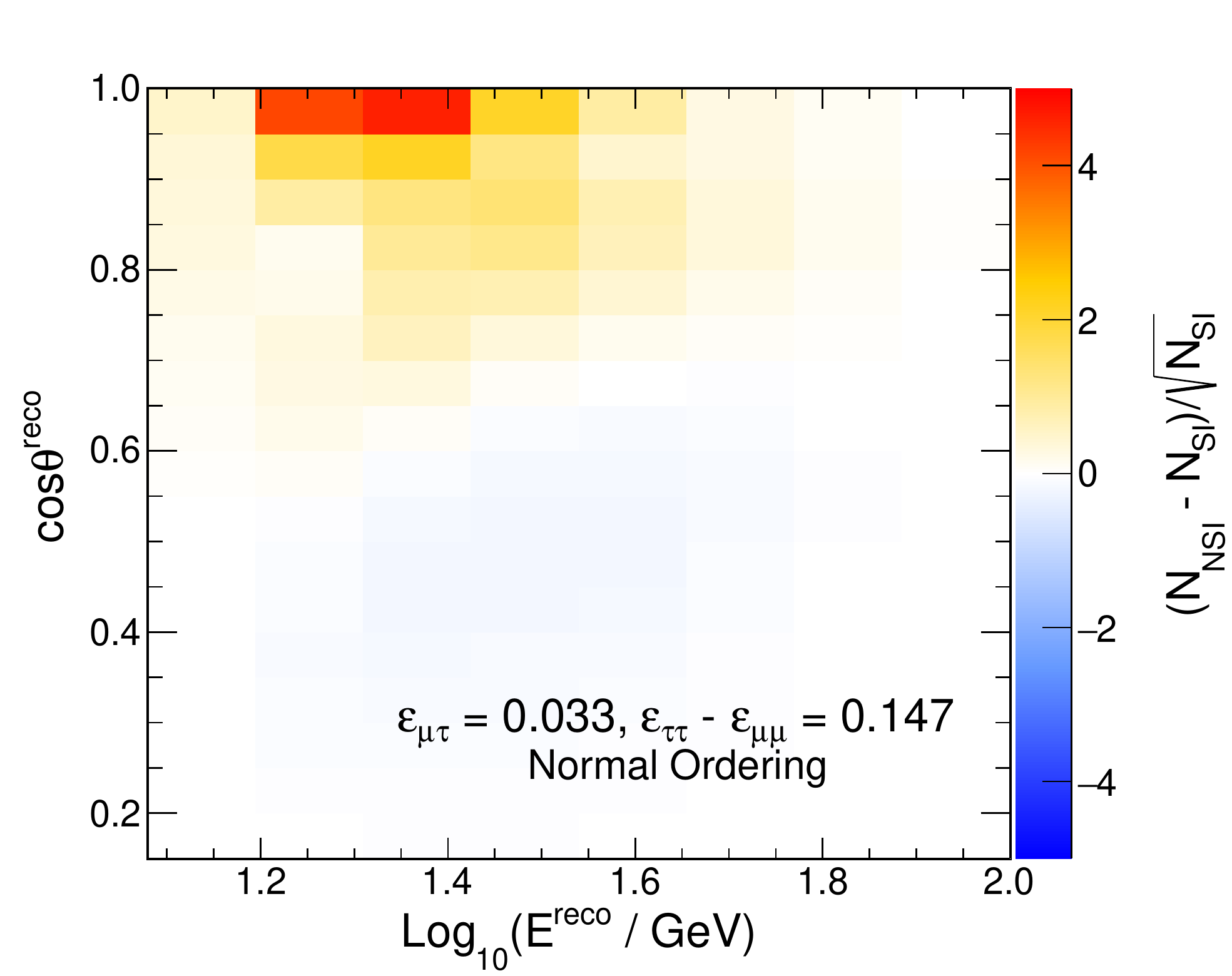}
\includegraphics[width=0.5\textwidth]{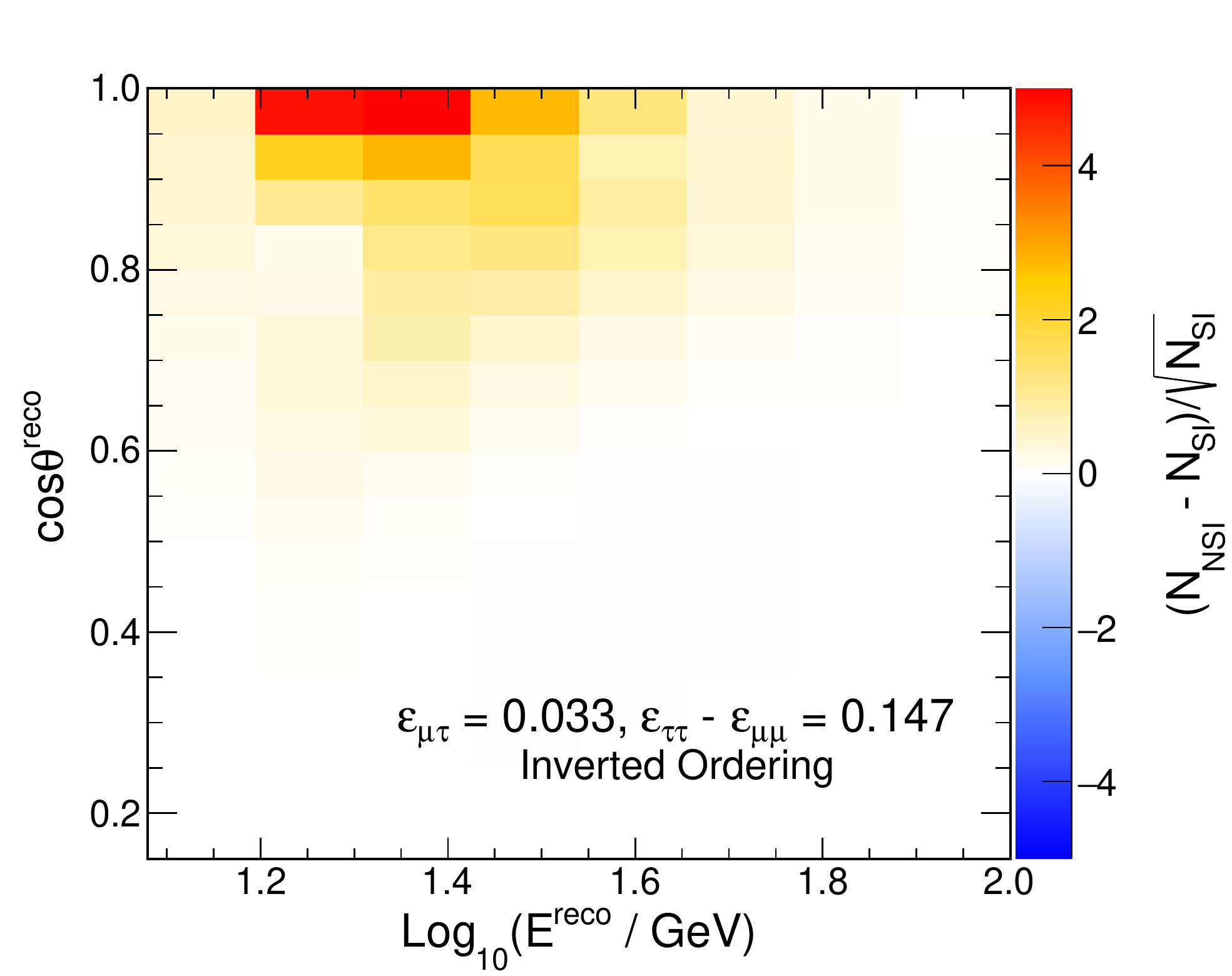}
\caption{Difference of number of events with and without NSI as a function of the reconstructed neutrino energy ($\log_{10}(E^{\rm reco}$)) and direction (cos$\,\theta^{\rm reco}$). Both panels correspond to a NSI test point, parameterised by $\epsab$ as quoted on the plots for normal (left) and inverted (right) orderings. 
}
\label{fig:statchi2antares}
\end{figure}

The selected events are distributed in a two-dimensional histogram of the reconstructed muon energy, $E^{\rm reco}$,  versus the cosine of the zenith angle, cos$\,\theta^{\rm reco}$, with a total of 136 bins. The muon energy spectrum is divided into eight bins: a first wide bin between 10$^{-0.3} = 5$~GeV and 10$^{1.2} = 15.8$~GeV, plus seven bins equally logarithmically spaced between 10$^{1.2}$ and 10$^{2}$~GeV. The reconstructed cos$\,\theta^{\rm reco}$ is divided into seventeen bins between 0.15 and 1.0, the latter value corresponding to vertically up-going events. The difference in the number of events expected with NSIs (assuming a trial set of values of model parameters, quoted on the plots) and without NSIs is shown in Figure~\ref{fig:statchi2antares}.

In order to extract the parameters that best describe the observed energy versus zenith angle distribution, the following test statistic is defined:

\begin{equation}
   -2 \, \text{log} \,\mathcal{L}  = 2\sum_{i,j} \Big (\mu_{i,j}(\bar{o}, \bar{s}) -n_{i,j}
 +  n_{i,j} \cdot \text{ln}\frac{n_{i,j}}{\mu_{i,j}(\bar{o}, \bar{s})}\Big ) \, + \sum_{k\in \{syst\}}\frac{(s_{k} - \hat{s_{k}})^{2}}{\sigma^{2}_{s_{k}}},  
\label{eqn:likelihood}
\end{equation}
\par\noindent where $n_{i,j}$ and $\mu_{i,j}(\bar{o}, \bar{s})$ are the number of measured and expected events in bin ($i,j$). $\{ \bar{o} \}$ and $\{ \bar{s} \}$ are the model parameters, which represent the oscillation and the additional nuisance parameters, respectively (see Table~\ref{tab:systs}). 
The second sum runs over the systematic uncertainties $s_{k}$, where $\hat{s}_k$ is the assumed prior on the $k$-th parameter and $\sigma^{2}_{s_{k}}$, its uncertainty~\cite{Gonzalez_Garcia_2004}.

The expected number of events for each reconstructed ($E^{\rm reco}$, cos$\,\theta^{\rm reco}$) bin are calculated by applying oscillation and systematic parameter weights to un-oscillated events in the Monte Carlo event sample.
Neutrino oscillation probabilities including NSI effects are calculated using the OscProb package \cite{OscProb} and the Earth density profile is approximated with 44 radial layers of constant density \cite{PREM}.


The impact of possible systematics effects on the analysis
is taken into account introducing eight nuisance parameters (see Table~\ref{tab:systs} of section~\ref{sec:results}).  
The first four ones ($\Delta m^{2}_{31}$, $\theta_{23}$, $\theta_{13}$ and  $\delta_{CP}$) are those neutrino oscillation parameters whose uncertainties can influence the result of our analysis, while 
the last four ones
($N_{\nu}$, $\Delta\gamma$, $\nu/\bar{\nu}$ 
and $\Delta M_{A}$) reflect the uncertainties in the atmospheric neutrino flux predictions,  the detector response and the neutrino interaction models.

The parameters of the standard oscillations are treated as follows. The solar mass splitting, $\Delta m^{2}_{21}$, and the corresponding mixing angle, $\theta_{12}$, are fixed to 7.4 $\times$ 10$^{-5}$~eV$^{2}$ and 33.62$^{\circ}$, respectively~\cite{RPP2020}, since they have a negligible influence on the atmospheric neutrino oscillations. 
On the other hand, the atmospheric oscillation parameters $\Delta m^{2}_{31}$ and $\theta_{23}$ are treated as nuisance parameters left free during the fit, i.e. without any priors.
The reactor angle, $\theta_{13}$, is assigned a Gaussian prior with a central value of 8.54$^{\circ}$ and an uncertainty of $\pm$0.28$^{\circ}$. Although no impact is found on the final results, the CP violating phase $\delta_{\rm CP} $ is fitted without prior. Separate fits are performed for the normal (NO) and inverted (IO) mass ordering hypotheses. All the above values are based on ref.~\cite{Esteban:2016qun}.

Our treatment of the systematics stemming from the uncertainties in the neutrino flux predictions, the detector response and the neutrino interaction models follows the one we applied in our latest neutrino oscillation analysis~\cite{Albert:2018mnz}.

For the atmospheric neutrino flux, the predictions by Honda \textit{et al.}~\cite{Honda:2015fha} at the Frejus site (latitude 45.1$^{\circ}$~N, close to that of ANTARES, 42.5$^{\circ}$~N), are used. The systematics coming from the uncertainties in these predictions are considered introducing two nuisance parameters: a global neutrino normalisation factor, $N_{\nu}$, and a possible deviation for the standard value of the spectral index, $\Delta\gamma$, that takes into account possible changes in the neutrino spectrum due to uncertainties in the primary cosmic ray spectrum. The global normalisation factor is treated as a free parameter without constraints, while the possible change in the spectral index is introduced with a 5\%-width Gaussian prior. Since only event rates are observed, these two parameters also take care of some other sources of systematics, as we explain below.

Uncertainties on the neutrino/anti-neutrino flux ratio, $\nu/\bar{\nu}$, and on the flux asymmetry between upgoing and horizontal neutrinos, $\nu_{\mathrm{up}}/\nu_{\mathrm{hor}}$, have also been taken into account. These uncertainties~\cite{Barr_2006} have been parametrised by the IceCube Collaboration~\cite{Aartsen:2017nmd} computing a correction on the number of expected events as a function of the neutrino energy, flavour, chirality, direction and the value of the uncertainty on the flux ratio. These two ratios considered are found to be strongly correlated~\cite{Albert:2018mnz}, thus a unique nuisance parameter is considered in the fit.

The systematics related to the detector response come from the uncertainties in the OM photon detection efficiencies and the water absorption length. Dedicated MC simulations have been generated with modified OM photon detection efficiencies and a modified water absorption length, assuming a variation of $\pm 10\%$ from the nominal value, but keeping the same wavelength dependence. 

The overall OM efficiency can be easily adjusted to the measured coincidence rates from $^{40}$K decays~\cite{K40Pap} which makes the chosen 10\% variations a conservative benchmark value, in line with early studies performed on ANTARES OMs~\cite{Amram:2001mi}. The water absorption length has been measured several times at the ANTARES site~\cite{Aguilar:2004nw}. The different measurements, taken at two different wavelengths, vary within about 10\%.
 
The change in the event rates, expressed as the ratio of the event rates with the modified MC simulation to the one from the nominal MC simulation, has been computed as a function of the MC neutrino energy and zenith angle for $\nu_{\mu}$ CC events, reconstructed as upgoing. While no zenith-dependent effect is seen, the energy response of the detector is affected by these variations. The resulting distributions have been fitted, in the energy range $10-10^3$\,GeV, with a function of the form:
\begin{equation}
f_\epsilon(\mathrm{E_T}) = A_\epsilon\cdot (\mathrm{E_T}/\mathrm{E_0})^{B_\epsilon},
\label{eq:systematic}
\end{equation}
where $\mathrm{E_T}$ is the MC true neutrino energy, $A_\epsilon$, $B_\epsilon$ are the two fitted parameters describing the effect of the modified OM photon detection efficiencies and $E_0 = 100$~GeV defines the reference energy for $A_\epsilon$. The $\pm 10\%$ variation in the 
OM efficiency translates into a $\pm 20\%$ in the event rate
at 100~GeV that decreases linearly in log($\mathrm{E_T}$) to $\pm 10\%$ at 1~TeV (see Fig.4 of ref.~\cite{Albert:2018mnz} for further details). The effect of the modified water absorption length is described by the same functional form of Eq.~\ref{eq:systematic} using $A_w$ and $B_w$ as the corresponding fit parameters. 

The effects of $A_\epsilon$ and $A_w$ are taken into account in the minimisation procedure by the global normalisation factor, $N_{\nu}$, while those of $B_\epsilon$ and $B_w$ are covered by the uncertainty of the prior on the spectral index, $\Delta\gamma$. Therefore, no new nuisance parameter has to be introduced to allow for systematics coming from OM efficiencies and water absorption.

The value of the atmospheric muon background contamination and its uncertainty have been obtained from the data itself following the method used in our oscillation paper~\cite{Albert:2018mnz} and described in the following. In the region of events with high $\chi^2$, which is dominated by atmospheric muons and excluded from the analysis, the data are fitted by exponential functions and extrapolated to the low $\chi^2$ region, where the neutrino signal is dominant and is used for the analysis. The integral in that region gives an estimate of the atmospheric muons. Different fit ranges in the high $\chi^2$ give different exponential fits. The final number of atmospheric muons is taken to be the mean and the uncertainty is estimated from the errors of the parameters of the fitted functions        
(for further details see section 5 of ref.~\cite{Albert:2018mnz} and  section 6.5.3 of ref.~\cite{Ilenia}). These values are subsequently used as the Gaussian prior mean and deviation in the minimisation procedure. The energy and direction distribution of the atmospheric muon background has been estimated directly from MC. In the present analysis, we fix the atmospheric muon contamination at the value obtained in the oscillation analysis~\cite{Albert:2018mnz}. Although this choice is motivated to ensure a stable fit procedure, it has been verified that this does not yield better constraints than an unconstrained muon normalisation.

Finally, a source of systematic uncertainty is the limited knowledge of the neutrino interaction model. In the energy range relevant for this analysis, the cross section is dominated by deep inelastic scattering (DIS) with smaller contributions from quasi elastic (QE) and resonant (RES) scattering. Uncertainties in the DIS cross section can be incorporated in the global flux normalisation factor $N_{\nu}$ as well as in the correction to the spectral index $\Delta\gamma$. For what concerns the QE and RES processes, dedicated studies have been performed~\cite{Ilenia} with gSeaGen~\cite{Distefano:2016bcw}, which uses GENIE~\cite{Andreopoulos:2015wxa} to model neutrino interactions. The dominant systematic error is found to be the one related to the axial mass for CC resonance neutrino production, $M_{A}$. Its default value is 1.12~$\pm$~0.22\,GeV~\cite{Andreopoulos:2015wxa}.  
By varying this parameter by $\pm 1\sigma$, a correction to the expected number of events as a function of the true neutrino energy has been computed. Although the correction is found to be small (see section 6.2.3 and Fig. 6.4 of ref.~\cite{Ilenia})
the parameterisation of this correction is used in the final fit.

\section{Results}
\label{sec:results}
The best-fit point is obtained by minimising $-2 \, \text{log} \,\mathcal{L}$ in Eq. \ref{eqn:likelihood}. Although the results are not expected to be strongly sensitive to the neutrino mass ordering (NMO), both possibilities are fitted separately. 
 The values of  the fitted parameters at the best-fit point for the NO and IO scenarios are shown in Table~\ref{tab:systs}. The flipping of the signs with the change of the NMO in the fits stems from the partial degeneracy between the sign of the $\epsmt$ and the NMO~\cite{Mocioiu:2014gua}.


 \begin{table}[ht]
\centering
\begin{tabular}{|c|c|c| c|}
\hline
 \textbf{Parameter}&\textbf{Prior} &\textbf{Best-fit value (NO)} &\textbf{Best-fit value (IO)}   \\
\hline
$\varepsilon_{\mu\tau}$  & none &$(-1.3^{+1.8}_{-2.0}) \times 10^{-3}$  & $(1.3^{+1.9}_{-1.8}) \times 10^{-3}$ \\
$\varepsilon_{\tau\tau} - \varepsilon_{\mu\mu}$ & none  & $(3.2_{-0.8}^{+1.4}) \times 10^{-2}$& $(-3.2_{-1.1}^{+1.9}) \times 10^{-2}$\\
$\Delta m^{2}_{31} [10^{-3} \rm eV^{2}]$& none & $3.0_{-0.6}^{+0.8}$&$-3.0^{+0.6}_{-0.8}$ \\
$\theta_{23} [^\circ]$ &none& $51\pm 9$ & $51\pm 9$\\
$\theta_{13} [^\circ]$ & 8.54 $\pm$ 0.28& 8.4 $\pm$ 0.3& 8.4 $\pm$ 0.3\\
$\delta_{\rm CP} [^\circ]$ &none&  $24_{-24}^{+340}$ & $176^{+180}_{-180}$\\
$N_{\nu} $ & none & 0.9 $\pm$ 0.1 & 0.9 $\pm$ 0.1 \\
$\Delta\gamma$ & 0.00 $\pm$ 0.05 & $-$0.02 $\pm$ 0.04 & $-$0.02 $\pm$ 0.04 \\
$\nu/\bar{\nu}$ $[\sigma]$ & 0 $\pm$ 1 & $1.0 \pm 0.6$& $1.0 \pm 0.6$\\
$\Delta \rm M^{\rm RES}_{A} [\sigma]$ & 0 $\pm$ 1 & 0.1 $\pm$ 1.0 & 0.1 $\pm$ 1.0\\
\hline
\end{tabular}
\caption{List of free parameters used in the maximisation of the likelihood ratio, their priors and the best-fit values for the case of NO and IO hypotheses.}
\label{tab:systs}
\end{table}


Concerning the nuisance parameters, the following features can be observed. The values of $\Delta m^2_{31}$ and  $\theta_{23}$ are in agreement within errors with our results for the standard oscillation analysis~\cite{Albert:2018mnz} and with the world best-fit values~\cite{RPP2020}. The value of $\theta_{13}$ is very close to its prior, indicating poor sensitivity to this parameter, which is to be expected since the $\nu_{\mu}$ survival probability depends on cos $\theta_{13}$, which is very close to one. 

The global normalisation factor, $N_{\nu}$, is 10\% lower than 1. This is in line with the result obtained in our standard oscillation analysis~\cite{Albert:2018mnz}, where a value 18\% lower than 1 was found. Both values are within the atmospheric neutrino flux uncertainties and are compatible with what has been reported by other analyses~\cite{Aartsen:2017nmd}. The high pull in $\nu/\bar{\nu}$ is similar to the one obtained in our standard oscillation analysis. In that analysis, the conclusion was that this parameter seems to compensate the low value of $N_{\nu}$, as could be seen fixing all the nuisance parameters but $N_{\nu}$ (see section 6.1 in Ref.~\cite{Albert:2018mnz}).  

The parameter $\Delta \rm M^{\rm RES}_{A}$ remains very close to its prior, indicating a low sensitivity of the fit to it. As explained in the previous section, this is due to the relative high energy threshold of ANTARES.


\begin{figure}[ht]
\centering
\includegraphics[width=\textwidth]{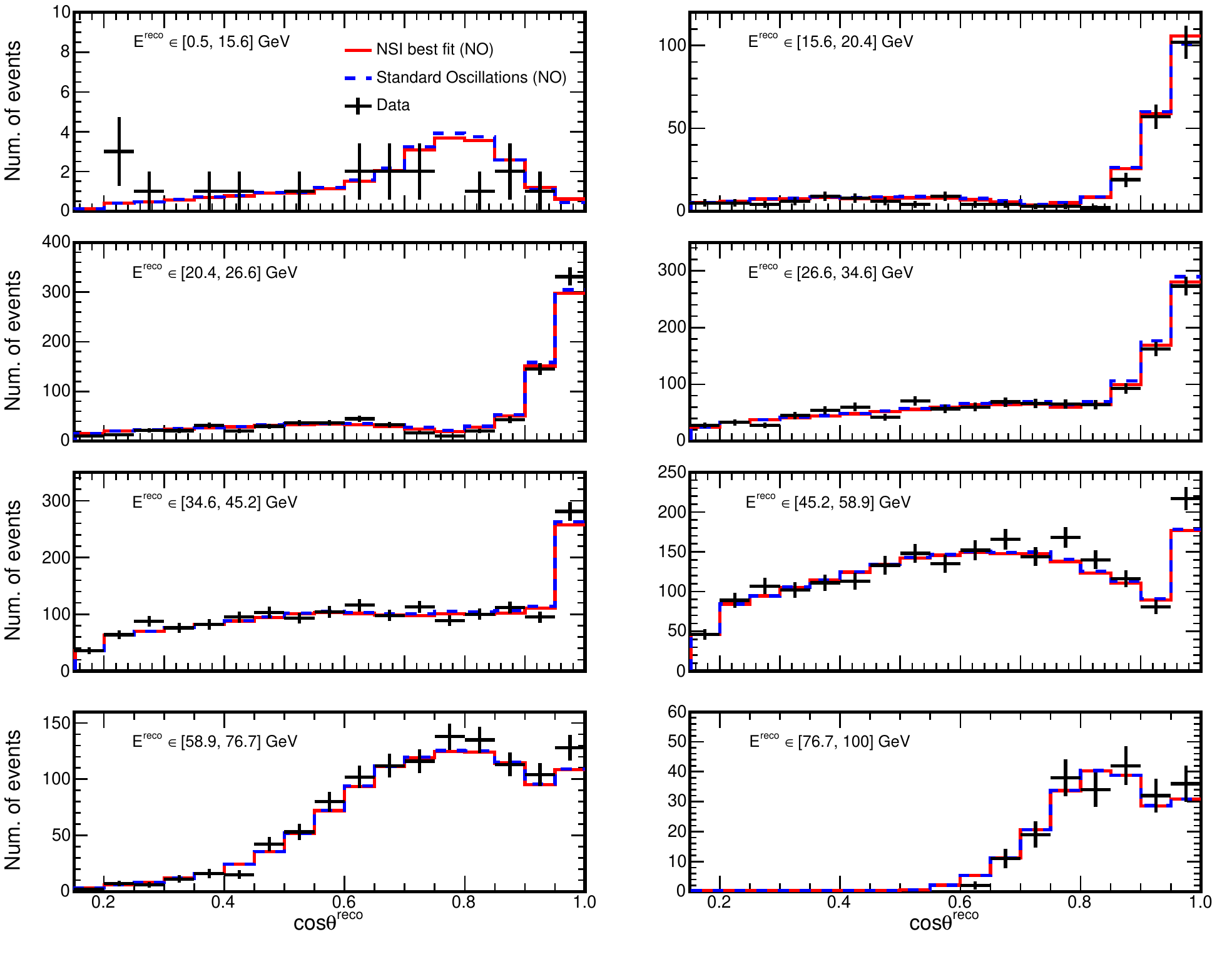}

\caption{Data (black points) and MC simulated events from this analysis, as a function of the reconstructed zenith angle, $cos\,\theta^{\rm reco}$, for the eight different energy, $E^{\rm reco}$ bins. The blue dashed histogram corresponds to the MC assuming standard oscillations and the (mostly overlapping) red histogram corresponds to the MC for the NSI case using the best-fit values obtained for normal ordering.  Note the difference in Y-axis scales in different panels.}
\label{fig:DataMCbestfitprojections}
\end{figure}

In Figure~\ref{fig:DataMCbestfitprojections}, the expected number of events using the best-fit values of Table~\ref{tab:systs} compared to the actual number of events observed by the experiment are shown. The number of events as a function of cos$\,\theta^{\rm reco}$ are displayed for the eight considered energy intervals, which are given in the figure in increasing order of energy from left to right and from top to bottom. The black crosses are the data with their statistical uncertainties and the red histograms are the expectations from the best-fit values obtained by maximising the log-likelihood in Eq.~\ref{eqn:likelihood}. The global minimum is found in the fit of NO. The dashed blue histogram is the expectation for the standard oscillation hypothesis.

\begin{figure}[ht]
\centering
\includegraphics[width=0.495\textwidth]{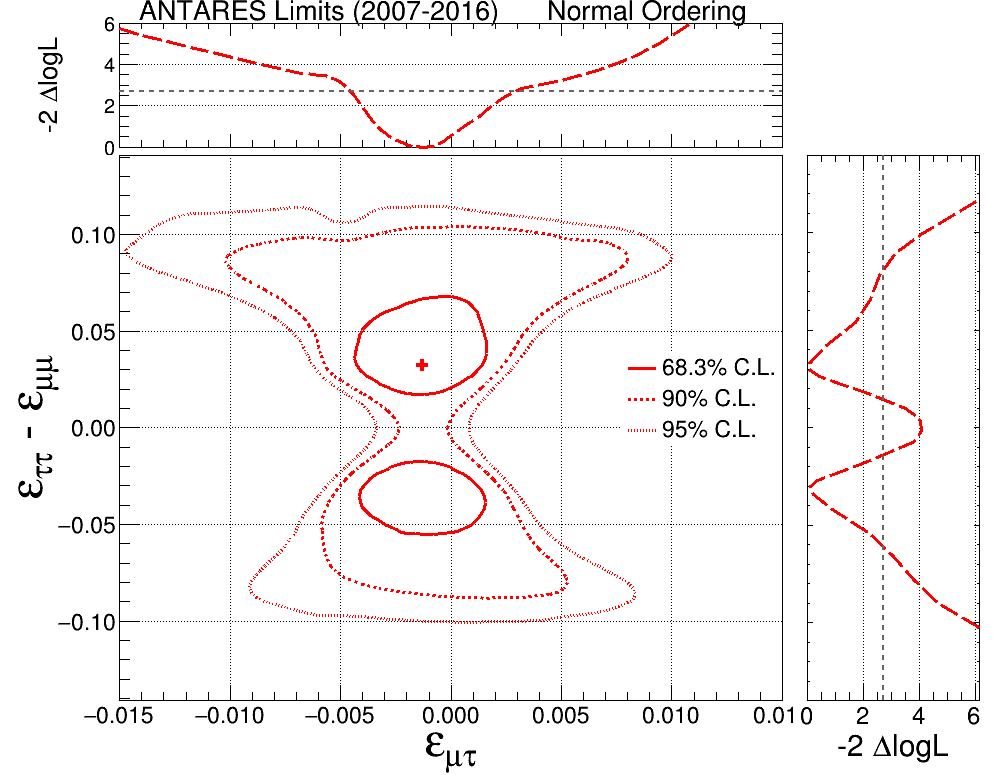}
\includegraphics[width=0.495\textwidth]{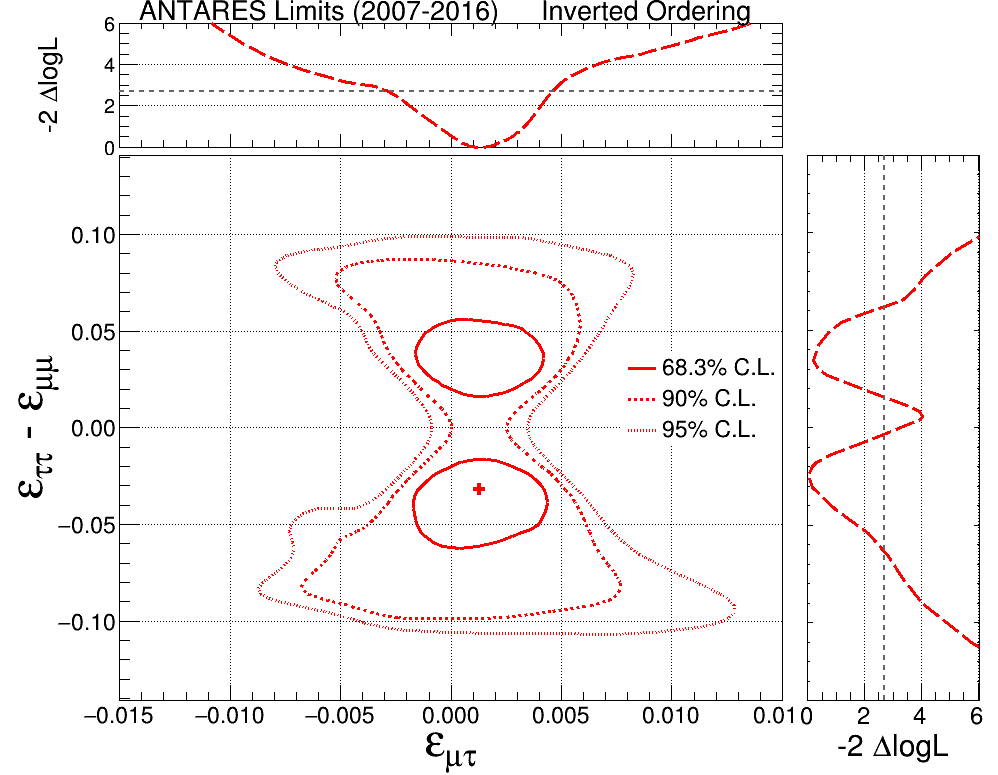}

\caption{Contours of 68.3\% (solid), 90\% (dashed) and 95\% (dotted) confidence level on the ($\varepsilon_{\mu\tau}, \, \varepsilon_{\tau\tau} - \varepsilon_{\mu\mu}$) plane, after 10 years of ANTARES livetime for the NO (left) and the IO (right) cases. The cross indicates the best-fit point obtained in both cases. The lateral plots on both panels show the 1D profile likelihood of the respective NSI parameters under study, when the other parameter is fitted over. The dashed straight lines in the lateral plots indicate the 90\% C.L. for one d.o.f.}
\label{fig:limitsantaresallquad}
\end{figure}

Exclusion contours in the ($\varepsilon_{\mu\tau}$, $\varepsilon_{\tau\tau} - \varepsilon_{\mu\mu}$) plane are drawn in the form of confidence level intervals for two degrees of freedom (d.o.f.). 
For each point of a grid in the ($\varepsilon_{\mu\tau}$, $\varepsilon_{\tau\tau} - \varepsilon_{\mu\mu}$) plane, the quantity $-2 \, \text{log} \,\mathcal{L}$ is minimised leaving free the rest of parameters. The differences between the values of $-2 \, \text{log} \,\mathcal{L}$ at the absolute minimum and at the grid points are used to build significance contours for two d.o.f., assuming that it obeys the $\chi^{2}$-distribution asymptotically~\cite{Wilks:1938dza}. In Figure~\ref{fig:limitsantaresallquad}, the resulting exclusion limits at various C.L. intervals are shown.

Finally, limits on $\varepsilon_{\mu\tau}$ and  $\varepsilon_{\tau\tau} - \varepsilon_{\mu\mu}$ are obtained by profiling over the other variable, as shown in the one-dimensional plots in the top and right panels of Figure~\ref{fig:limitsantaresallquad}. The 90\% C.L. bounds are:

\begin{equation}
\begin{array}{rcl}
         -4.7 \times 10^{-3} < \varepsilon_{\mu\tau} < 2.9 \times 10^{-3} \quad  \qquad \qquad \qquad  \,\, &(\text{NO}),\\ \\
         
     -6.1\times 10^{-2} < \varepsilon_{\tau\tau} -  \varepsilon_{\mu\mu}  < -1.4 \times 10^{-2} \quad  \qquad \qquad \qquad  \\
     
     \cup \quad  \qquad \qquad \qquad \quad \qquad \qquad \quad \quad &(\text{NO}), \\ 
     
      1.4\times 10^{-2} < \varepsilon_{\tau\tau} -  \varepsilon_{\mu\mu} < 8.1\times 10^{-2} \quad  \quad \quad \quad \quad \,\,\,\, \\  \\

\\ 

         -2.9 \times 10^{-3} < \varepsilon_{\mu\tau} < 4.7 \times 10^{-3} \quad  \qquad \qquad \qquad  \,\, &(\text{IO}),\\ \\
         
     -6.4\times 10^{-2} < \varepsilon_{\tau\tau} -  \varepsilon_{\mu\mu}  < -0.4 \times 10^{-2} \quad  \qquad \qquad \quad \\
     
     \cup \quad  \qquad \qquad \qquad \quad \qquad \qquad \quad \quad  &(\text{IO}). \\
     
     1.4\times 10^{-2} < \varepsilon_{\tau\tau} -  \varepsilon_{\mu\mu}  < 6.4\times 10^{-2}  \quad  \quad \quad \quad \quad  \,\,\,\,  \\
\end{array}
\label{eq:antaresmtttlimitsummary}
\end{equation}

We can summarise the results as follows. The hypothesis of standard interactions, i.e. the compatibility with zero of the combined fit result, is disfavoured with a significance of  1.7$\sigma$ (1.6$\sigma$) for the normal (inverted) mass ordering scenario. However, the 90\% C.L. contours do contain the null value. The difference from zero is mainly due to the $\varepsilon_{\tau\tau}  - \varepsilon_{\mu\mu}$ coefficient. Indeed, while the value obtained for $\varepsilon_{\mu\tau}$ profiling over the other coefficient is compatible with zero within its 68\% C.L.
interval, the 90\% C.L. intervals for $\varepsilon_{\tau\tau}  - \varepsilon_{\mu\mu}$ do not include the null value. However, the 95\% C.L. interval for $\varepsilon_{\tau\tau}  - \varepsilon_{\mu\mu}$ does contain zero. Therefore, we conclude that the results do not show clear evidence of deviations from standard interactions



The limits obtained in this analysis are among the most stringent published to date for these NSI parameters, in particular for $\varepsilon_{\mu\tau}$. In Table~\ref{tab:comparison}, the 90\% CL limits on $\varepsilon_{\mu\tau}$ and $\varepsilon_{\tau\tau} - \varepsilon_{\mu\mu}$ provided by atmospheric neutrino experiments are summarised.  

When comparing the results on Table~\ref{tab:comparison} the following caveats should be considered. Some results were obtained assuming $\varepsilon_{\mu\tau}$ to be real, as in our analysis, while some others assume it to be complex and fit both the modulus and the phase. For the latter, the number quoted is the 90\% CL upper limit on the modulus. The two cases are indicated by notes (a) and (b) in the fourth column of the table. Although the sensitivity to the phase is not high in general for this sort of experiments and the limits tend to weaken with respect to the assumption of a real $\varepsilon_{\mu\tau}$, a complex $\varepsilon_{\mu\tau}$ is the most general hypothesis.

\begin{table}[H]
\centering
\begin{tabular}{|c|c|c|c|c|}
\hline
 & & & &\\
 Experiment & \makecell{Reconstructed\\ Energy range \\ (GeV)} & \makecell{$\varepsilon_{\mu\tau} $ or $|\varepsilon_{\mu\tau}|$ \\$(\times 10^{-3})$}  & 
  \makecell{$\varepsilon_{\tau\tau} -  \varepsilon_{\mu\mu}$ \\ $(\times 10^{-2})$} & \\
 & & & &\\
\hline
$\text{ANTARES}$ $(\text{this work})$ & $15.8 - 100$ &
$[-4.7, 2.9]$ &  
\makecell{$[ -6.1, -1.4]$ \\ $\cup$ \\ $[1.4, 8.1]$} 
& (a)(d)\\
 & & & &\\
$\text{IceCube 2022}$ \cite{IceCube:2022ubv} & 
$5 \times 10^{2} - 10^{4}$ &
$[-4.1, 3.1]$ &   & (a)(d)\\
 & & & &\\
 & & & &\\
$\text{IC DeepCore 2021}$ \cite{IceCubeCollaboration:2021euf} & $5.6 - 100$ & \makecell {$[-16.5, 13.0]$ \\ \\ $23.2$ }  &  $[-4.1, 4.2]$  & 
\makecell {(a)(c) \\ \\ (b)(c)} \\
 & & & &\\
 & & & &\\
$\text{IC DeepCore 2020 (Pub)}$ \cite{Demidov:2019okm} & $5.6 - 56$ &
$[-23, 16]$ & $[-5.5, 5.6]$  & (a)(c)\\
  & & & &\\
  & & & &\\  
$\text{IC DeepCore 2018}$  \cite{Aartsen:2017xtt} (\cite{IceCubeCollaboration:2021euf}) & $5.6 - 56$&
\makecell{ $[-6.7, 8.1]$ \\ \\ $[-20, 24]$ } &   & 
\makecell{(a)(d) \\ \\ (a)(c)} \\
 & & & &\\
 & & & &\\
$\text{IceCube 2017 (Pub)}$  \cite{Salvado:2016uqu} & $3 \times 10^2 - 2 \times 10^4$  &
$[-6.0, 5.4]$ &   & (a)(c)\\
  & & & &\\
$\text{Super-Kamiokande}$  \cite{Mitsuka:2011ty} & $\gtrsim 1 - \lesssim 100$ &
$11$ & $[-4.9, 4.9]$  & (b)(d)\\
  & & & &\\
\hline 
\end{tabular}

\caption{Summary of the 90\% CL upper limits on the 
NSI parameters $\varepsilon_{\mu\tau} $ (or $|\varepsilon_{\mu\tau}|$) and $\varepsilon_{\tau\tau} -  \varepsilon_{\mu\mu}$ from different atmospheric neutrino experiments. The intervals indicate the regions not excluded by the corresponding analysis. For $|\varepsilon_{\mu\tau}|$ the single upper limit is given.
The label "Pub" indicates that the public data have been analysed by researchers outside the collaboration.  
The conditions applied are as follows:
(a) assuming real $\varepsilon_{\mu\tau}$, (b) modulus of complex $\varepsilon_{\mu\tau}$, (c)~for effective NSI couplings and (d) for d-quark NSI couplings. All the results are given for normal ordering. 
The approximate neutrino energy range used in each analysis is also shown.
See the main text for explanations. 
}
\label{tab:comparison}
\end{table}

Some results are given using NSI couplings to down-quarks while some others are given for effective NSI couplings to electrons, protons and neutrons. The limits of the former should be multiplied by a factor $\approx$3 to compare to the latter. Limits obtained with the effective coupling and down-quarks couplings are denoted as (c) and (d), respectively. Whenever the authors themselves provide the translation from one case to the other we list both intervals (e.g. for IC DeepCore 2018~\cite{Aartsen:2017xtt} translated in ref.~\cite{IceCubeCollaboration:2021euf}). All the limits are given for normal ordering. The table also shows the range of the reconstructed energy of the neutrino sample for each analysis. 

In addition to the results based on atmospheric neutrino data, there is a large variety of bounds to NSI parameters from other analyses, some of which are collected in ref.~\cite{Farzan:2017xzy}.
Bounds extracted from the data of one or more long baseline accelerator experiments (NOMAD, MINOS, T2K, NO$\nu$A) are given, for instance, in refs.~\cite{Biggio:2009nt, MINOS:2013hmj, Denton:2020uda}. Bounds based on the data from the COHERENT experiment are given in refs.~\cite{Denton:2018xmq, Giunti:2019xpr}.
Bounds extracted from global fits to the oscillation data plus the results of the COHERENT experiment are given in refs.\cite{Esteban:2018ppq, Esteban:2019lfo, Coloma:2019mbs}. Notice that not all of these bounds can be compared right away to those obtained from atmospheric neutrinos and some of them (e.g. those from the CE$\nu$NS data) are model dependent.


\section{Conclusions}
\label{sec:conclusion}

The search for neutral current non-standard interactions in neutrino propagation using 10 years of ANTARES data is presented. We do not find clear evidence of deviations from standard interactions. For normal (inverted) neutrino
mass ordering, the combined fit of both NSI coefficients yields a value of 1.7$\sigma$ (1.6$\sigma$) away from the null result. However, the 68\% and 95\% confidence level intervals for $\varepsilon_{\mu\tau}$ and $\varepsilon_{\tau\tau} - \varepsilon_{\mu\mu}$, respectively, contain the null point. 
The resulting bounds of ANTARES obtained from this analysis are among the most restrictive to date. 

ANTARES has recorded additional data from 2017 to 2022, which represents a $\sim$50\% increase with respect to that used in the present analysis and are currently being analysed. Furthermore, the next generation of neutrino telescopes in the Mediterranean Sea, KM3NeT-ORCA and -ARCA, now under deployment~\cite{KhanChowdhury:2020xev, KhanChowdhury:2020qqu, Kalaczy_ski_2021, Fermani:2019vac}, are expected to further improve these limits.

\section{Acknowledgements}
The authors acknowledge the financial support of the funding agencies:
Centre National de la Recherche Scientifique (CNRS), Commissariat \`a
l'\'ener\-gie atomique et aux \'energies alternatives (CEA),
Commission Europ\'eenne (FEDER fund and Marie Curie Program),
Institut Universitaire de France (IUF), LabEx UnivEarthS (ANR-10-LABX-0023 and ANR-18-IDEX-0001),
R\'egion \^Ile-de-France (DIM-ACAV), R\'egion
Alsace (contrat CPER), R\'egion Provence-Alpes-C\^ote d'Azur,
D\'e\-par\-tement du Var and Ville de La
Seyne-sur-Mer, France;
Bundesministerium f\"ur Bildung und Forschung
(BMBF), Germany; 
Istituto Nazionale di Fisica Nucleare (INFN), Italy;
Nederlandse organisatie voor Wetenschappelijk Onderzoek (NWO), the Netherlands;
Council of the President of the Russian Federation for young
scientists and leading scientific schools supporting grants, Russia;
Executive Unit for Financing Higher Education, Research, Development and Innovation (UEFISCDI), Romania;
Ministerio de Ciencia, Innovaci\'{o}n, Investigaci\'{o}n y
Universidades (MCIU): Programa Estatal de Generaci\'{o}n de
Conocimiento (refs. PGC2018-096663-B-C41, -A-C42, -B-C43, -B-C44)
(MCIU/FEDER), Generalitat Valenciana: Prometeo (PROMETEO/2020/019),
Grisol\'{i}a (refs. GRISOLIA /2018/119, /2021/192) and GenT
(refs. CIDEGENT/2018/034, /2019/043, /2020/049, /2021/023) programs, Junta de
Andaluc\'{i}a (ref. A-FQM-053-UGR18), La Caixa Foundation (ref. LCF/BQ/IN17/11620019), EU: MSC program (ref. 101025085), Spain;
Ministry of Higher Education, Scientific Research and Innovation, Morocco, and the Arab Fund for Economic and Social Development, Kuwait.
We also acknowledge the technical support of Ifremer, AIM and Foselev Marine
for the sea operation and the CC-IN2P3 for the computing facilities.


\bibliography{ref}

\end{document}